\documentclass[aps,prd,onecolumn,superscriptaddress,showpacs,floatfix]{revtex4-2}%

\usepackage{graphicx}
\usepackage{tabularx}
\usepackage{bm}
\usepackage{latexsym}
\usepackage{epsf}
\usepackage{rotating}
\usepackage{epsfig,graphics,rotate,color}
\usepackage{wrapfig}
\usepackage{amssymb}
\usepackage{stmaryrd}
\usepackage{amsmath}
\usepackage{amsfonts}
\usepackage{comment}
\usepackage{subfigure}
\usepackage{array,hhline,dcolumn}%
\usepackage[normalem]{ulem}
\usepackage{color}
\usepackage{wasysym}
\usepackage[colorlinks=true,allcolors=blue]{hyperref}

\begin{document}
\bibliographystyle{apsrev4-1}

\title{The Width of an Electron-Capture Neutrino Wave Packet 
}

\author{B.J.P. Jones}
\affiliation{
University of Texas at Arlington, Arlington, Texas 76019, USA
}

 \author{E. Marzec}
\affiliation{
 University of Michigan, Ann Arbor, Michigan 48109, USA
 }

 \author{J. Spitz}
 \affiliation{
 University of Michigan, Ann Arbor, Michigan 48109, USA
 }

\begin{abstract}
We discuss the size of an electron neutrino wave packet emerging from an electron capture decay using the formalism of open quantum systems. This quantitative result is based on methodology that we have previously used to predict the width of an electron antineutrino wave packet from a beta-decaying nucleus, adapted for the different initial states in electron capture and beta decay. These predictions are now becoming relevant to experimental probes of the neutrino width, for example indirectly through precision spectroscopic studies of the nuclear recoil.  We provide a translation between a recent Beryllium Electron capture in Superconducting Tunnel junctions Experiment (BeEST) measurement of the daughter nuclear recoil spectrum from $\mathrm{^{7}Be}$ electron capture decay ($e^- + \mathrm{^{7}Be}\rightarrow\mathrm{^{7}Li+\nu_e}$) and a constraint on the outgoing neutrino width. According to our
analysis, the direct limit on the neutrino wave packet width from electron capture decay using the recent BeEST result should map to $\sigma_{\nu,x}>6.2\,\mathrm{pm}$.
We determine that a hard upper limit on the wave packet width exists at $\sigma_{\nu,x}\sim2.7\,\mathrm{nm}$, based on the relative momentum between the electron and nucleus. However, we find that the localization is most likely driven by the embedding of the decaying atom in the tantalum crystal lattice resulting in a width scale of approximately $\sigma_{\nu,x}\sim10$~pm, motivated by data from X-ray diffraction and  M$\ddot{\mathrm{o}}$ssbauer spectroscopy, which lies intriguingly close to the current bound.




\end{abstract} \maketitle

\section{Introduction}

The coherent width of a neutrino wave packet is a quantum mechanical
quantity that is expected to impact neutrino oscillation phenomenology
at long baselines \cite{akhmedov2009paradoxes,akhmedov2022damping,arguelles2023impact,daya2017study,giunti1998coherence,giunti2002neutrino,Marzec:2022mcz},
through coherence loss between interfering mass eigenstates. This
width is an emergent property, which is predictable from first principles
\cite{jones2023width,jones2015dynamical}, with dependencies on the
neutrino energy, initial flavor and production process. A recent paper
from the BeEST collaboration \cite{Smolsky:2024uby} aims to set
direct limits on the size of the neutrino wave packet emerging from electron capture ($e^- + \mathrm{^{7}Be}\rightarrow\mathrm{^{7}Li+\nu_e}$) via precision spectroscopy
of the entangled daughter recoil. The extremely precise energy resolution of
superconducting tunnel junctions make them the ideal sensors for this
type of measurement \cite{leach2022beest,kurakado1982possibility},
and since this is the first reported study of this kind, it represents an exciting advance in this area. 

While it is not expected that neutrino wave packet effects impacting neutrino interaction or oscillation observables in near-term neutrino experiments~\cite{jones2015dynamical,Marzec:2022mcz}, the prospect of indirect sensitivity to the neutrino wave packet width based on measurements of the recoiling daughter nucleus, as in BeEST, motivates predictions for this quantity. Our previous work~\cite{jones2023width} presents an  explicit calculation of the wave packet width of an antineutrino emerging from nuclear beta decay. We emphasize in that paper the role of the initial state delocalization scale in determining the wave packet width. In Section~\ref{predict_width}, we apply this methodology to the case of electron capture and produce a prediction for the neutrino width.  We further provide a translation between the observable nuclear recoil spectrum (peak width) in BeEST, or any other recoil-based precision spectroscopic probe, and the neutrino width itself in Section~\ref{map_recoil_neutrino}, and briefly comment on the possibility of neutrino wave packet effects as a possible explanation of the ``gallium anomalies"~\cite{abdurashitov1996russian,anselmann1995first,bahcall1997gallium,barinov2022results} in Section~\ref{gallium_anomaly}.



\section{Predicted width of an electron capture neutrino}
\label{predict_width}

The initial and final states in beta decay and electron capture are quite different:
in beta decay, a single nucleon that is localized within a nucleus
transforms via weak decay into a final state consisting of an entangled
system of electron, electron antineutrino, and nuclear recoil. In electron capture,
two initial particles, an atomic electron and parent nucleus, interact to yield a two particle final state of neutrino and daughter nucleus.
There is a relative momentum uncertainty between the two initial-state
particles in electron capture that corresponds to a distance uncertainty
on atomic scales. Our prescription~\cite{jones2023width} can be applied to electron capture,
but it must account for both initial state particles and conserve
energy and momentum in the process.

As a brief summary, our proposed prescription for calculating neutrino
wave packet widths is as follows:
\begin{enumerate}
\item Write down the initial state, accounting for as many entangled microscopic
degrees of freedom as necessary. If the recipe is followed sufficiently
carefully, there will be no ambiguity about which degrees of freedom
/ distance scales ultimately act to determine the neutrino wave packet
width. Any extra recursive delocalization ultimately disappears when
forming the reduced density matrix;
\item Conserve energy and momentum in the decay to form the final state
density matrix for the fully entangled system;
\item Trace out everything that is not the neutrino, to form the neutrino
reduced density matrix. This will account for the localizing influence
of everything except the neutrino that is entangled with it;
\item The full neutrino oscillation phenomenology can be obtained from the
neutrino reduced density matrix. The quantity playing the role of
the coherent wave packet width in this object is the off-diagonal
width of the position-space density matrix.
\end{enumerate}
We now proceed to follow this recipe for electron capture.

\medskip{}

We will consider an initial state comprised of an electron ($e$) and a nucleus ($N$),
with momentum-space wave function
\begin{equation}
\Phi_{i}=\int d^{3}P\int d^{3}p~ \Psi(\vec{P})\psi(\vec{p})|N(\vec{p}_{N})\rangle\varotimes|e(\vec{p}_{e})\rangle.
\end{equation}
Here the coordinates chosen are the sum and difference of the momenta,
\begin{equation}
p=p_{e/\nu}-p_{N},\quad P=p_{e/\nu}+p_{N}.
\end{equation}

A note is in order about this coordinate choice. For the initial state,
an alternative coordinate choice would be the center-of-mass and reduced-mass
coordinates
\begin{equation}
q=\mu\left(\frac{p_{e}}{m_{e}}-\frac{p_{N}}{m_{N}}\right),\quad Q=p_{e}+p_{N}.
\end{equation}
This coordinate system has the convenient feature that variable $q$
satisfies the Schrödinger equation with reduced mass $\mu$, and it is the
one that is typically used to find the electron orbitals of an atom
in the center of mass frame. On the other hand, since in the final
state the neutrino is relativistic, these coordinates eventually become
a burden. It is important to observe, however, that in the center
of mass frame where $p_{e}=-p_{N}$, there is an equivalence $q=p$
and $Q=P$. As such, we can assert that the center of mass wave function $\phi(p)$
will be well approximated by $\phi(q)$, as long as we restrict our
calculations to frames where the neutrino carries far more of the
energy than the nucleus. The frames we are interested in will always
satisfy this requirement.

The electron capture process is expressed in terms of initial $N_{i}$
and final $N_{f}$ nuclei as

\begin{equation}
N_{i}+e\rightarrow N_{f}+\nu_{e}.
\end{equation}

For a sufficiently long-lived parent, we can assume energy and momentum
to be conserved in the decay, for practical purposes. For a sufficiently
short lifetime, an intrinsic energy uncertainty is also required by
the energy-time uncertainty principle, but this will be a small contribution
for the $^{7}$Be system. Thus we can schematically consider this
decay as being mediated by the following quantum operator, which conserves
momentum in the decay, and also that energy will be conserved in the
decay amplitude,
\begin{equation}
{\cal O}=\int d^{3}p_{i}d^{3}p_{e}d^{3}p_{f}d^{3}p_{\nu}\,a_{N_{i}}^{p_{i}}a_{e}^{p_{e}}a_{N_{f}}^{p_{f}\dagger}a_{\nu}^{p_{\nu}\dagger}\delta^{3}(p_{i}+p_{e}-p_{f}-p_{\nu}).\label{eq:Operator}
\end{equation}
Operator \ref{eq:Operator} conserves the center of mass momentum
of the whole system, but changes the relative momentum between the
lepton and nucleus. Thus the final state system wave function $\Phi_{f}$
must be expressible as
\begin{equation}
\Phi_{f}=\int d^{3}P\int d^{3}p~\Psi(\vec{P})\left[\phi(\vec{p})|N_{f}(\vec{p}_{N_{f}})\rangle\varotimes|\nu(\vec{p}_{\nu})\rangle\right].
\end{equation}
Where $\Psi$ is the same as it was before (the total delocalization
wave function of the center of mass in momentum space, which we assume
to be a Gaussian with width $\sigma_{P}=1/2\sigma_{X}$) and $\phi$
is a new center of mass wave function that is derivable from $\psi$
through kinematic considerations. 

For capture from the K shell the relative momentum wave function is
spherically symmetric $\psi(\vec{p})=\psi(p)$. Assuming nothing in
the initial state to be spin-polarized, the final state will also
be spherically symmetric in the center of mass frame $\phi(\vec{p})=\phi(p)$.
Since the nuclei are both non-relativistic, energy conservation requires
that
\begin{equation}
\left(m_{N_{i}}+\frac{p_{N_{i}}^{2}}{2m_{N_{i}}}\right)+\sqrt{p_{e}^{2}+m_{e}^{2}}=\left(m_{N_{f}}+\frac{p_{N_{f}}^{2}}{2m_{N_{f}}}\right)+\left|p_{\nu}\right|.
\end{equation}

If we consider this expression in the center of mass frame then we
may write
\begin{equation}
p_{e}=-p_{N_{i}}=p_{i},\quad p_{\nu}=-p_{N_{f}}=p_{f},
\end{equation}
\begin{equation}
p_{i}^{2}\left(\frac{1}{2m_{e}}+\frac{1}{2m_{N_{i}}}\right)+M=\frac{p_{f}^{2}}{2m_{N_{f}}}+|p_{f}|.\label{eq:Quadratic}
\end{equation}
Here $M=m_{N_{i}}+m_{e}-m_{N_{f}}$. The nucleus is much heavier
than both the electron mass and the relevant kinetic energies in the
problem, so we can to a reasonable degree of approximation neglect
the $1/m_{N}$ terms, to give the following relation between initial
and final center of mass momenta,
\begin{equation}
|p_{f}|=M+\frac{p_i^{2}}{2m_{e}}.\label{eq:EnergyMatch-1}
\end{equation}
This approximation is equivalent to the statement that in the center
of mass frame where momentum in the decay is shared equally, the electron
or neutrino must carry essentially all of the kinetic energy in the
initial or final state, respectively. A more accurate relation can
be obtained using the quadratic solution to Eq.~\ref{eq:Quadratic}
and following the steps below, adding complexity but with negligible
impact on the final answer.

We consider here only the K shell electron capture, so the relevant initial wave function
is approximately the 1S electron wave function with atomic number
$Z=4$,
\begin{equation}
\psi(\vec{x})=\frac{Z^{3/2}}{a_{0}^{3/2}}\frac{1}{\sqrt{\pi}}e^{-Zr/a_{0}},
\end{equation}
where $a_0$ is the Bohr radius. To move into momentum space we Fourier transform in 3D:
\begin{equation}
\psi(\vec{p})=\int d^{3}xe^{i\vec{p}\cdot\vec{x}}\psi(\vec{x})=\int d\cos\theta\,d\phi\,drr^{2}e^{i\vec{p}\cdot\vec{x}}\psi(\vec{x}).
\end{equation}
Choosing coordinates for the $\vec{x}$ integral such that $\vec{p}$
points along $\theta=0$,
\begin{eqnarray}
\psi(\vec{p})=2\pi\int d\cos\theta\,drr^{2}e^{ipr\cos\theta}\psi(r)
\\
=\frac{8}{\sqrt{\pi}}\frac{Z^{3/2}}{a_{0}^{3/2}}\frac{a_{0}^{3}Z}{\left(a_{0}^{2}p^{2}+Z^{2}\right)^{2}} \equiv{\cal A}\left(1+\frac{p_{i}^{2}}{\left(Z/a_{0}\right)^{2}}\right)^{-2}.\label{eq:Fourier1S}
\end{eqnarray}

We have collected several constants into a normalization factor ${\cal A}$
that we can restore later if needed. To obtain $\phi$ in terms of
$\psi$, consider that each initial momentum magnitude $p_{i}$
maps to a unique final momentum magnitude $p_{f}$. Thus their wave
function amplitudes should also map, as
\begin{equation}
\phi(p_{f})=f[\psi(p_{i}\left[p_{f}\right]),p_{f}].
\end{equation}
The precise shape for this final state wave function depends
on whether we consider the 1D or 3D problem, but for our intended degree of precision here it suffices to use an example wave function with
the correct mean and width. We will thus use our trusty workhorse,
the Gaussian wave packet,
\begin{equation}
\phi(p_{f})={\cal B}\exp\left[-\frac{\left(p-p_{rel}\right)^{2}}{4\sigma_{rel}^{2}}\right]~,
\end{equation}
where $p_{rel}$ and $\sigma_{rel}$ represent the relative momentum and width, respectively. We estimate $p_{rel}$ and $\sigma_{rel}$ by noting that peak of
Eq.~\ref{eq:Fourier1S} occurs at $p_{i}^{mean}=0$ with its half width at half maximum (HWHM) at
($p_{i}^{HWHM})^2=\left(\sqrt{2}-1\right)(Z/a_{0})^2.$ This suggests we
might as a reasonable estimate take $\phi$ to have peak and HWHM
at the kinematically related points at the corresponding neutrino
momenta (with an extra factor of $\frac{1}{2}$ to account for the fact the kinematically related peak is one sided), 
\begin{equation}
p_{rel}=M,\quad\quad\quad\sigma_{rel}\sim\frac{1}{2\sqrt{2\mathrm{ln}2}}\left[M+\frac{p_{i}^{2}}{2m_{e}}\right]_{0}^{p_{i}^{HWHM}}=\frac{\zeta Z^{2}}{2 a_{0}^{2}m_{e}}.\label{eq:Quantities}
\end{equation}
With $\zeta=\frac{\sqrt{2}-1}{2\sqrt{2\ln2}}\sim0.18$. To find the
quantity representing neutrino wave packet width using the methodology
of Ref. \cite{jones2023width}, we need to calculate the off-diagonal
width of the reduced neutrino density matrix. This full-system density
matrix is defined as
\begin{equation}
\rho=|\Phi_{f}\rangle\langle\Phi_{f}|.
\end{equation}
The neutrino reduced density matrix, from which all oscillation phenomenology
can be calculated, is
\begin{equation}
\rho_{\nu}=\int dp_{N}\langle p_{N}|\rho|p_{N}\rangle.
\end{equation}
Writing this out with all its ingredients, where subscripts 1,2 refer to the variables that appear on the left and right of the density matrix, and $|\nu(p)\rangle$ is a neutrino state vector with momentum $p$,
\begin{equation}
\rho_{\nu}=\int d^{3}P_{1}\tilde{\Psi}(\vec{P_{1}})\left[\int d^{3}p_{1}\,\tilde{\psi_{f}}(\vec{p}_{1})\right]\int d^{3}P_{2}\tilde{\Psi}(\vec{P}_{2})\left[\int d^{3}p_{2}\,\tilde{\psi_{f}}(\vec{p}_{2})\right]\delta(p_{N1}-p_{N2})\,|\nu(\vec{p}_{\nu1})\rangle\langle\nu(\vec{p}_{\nu2})|
\end{equation}

We will carry all the Gaussian normalization factors forward in an arbitrary
constant ${\cal N}$, which will cancel when we calculate the quantities
of interest. 
Considering the problem in one dimension, and evaluating the intermediate integrals to find the reduced density operator, we obtain
\begin{equation}
\rho={\cal N}\int dp_{\nu1}dp_{\nu2}\exp\left[-\frac{1}{8}\left(\frac{1}{\sigma_{P}^{2}}+\frac{1}{\sigma_{rel}^{2}}\right)\left(p_{\nu1}-p_{\nu2}\right)^{2}-\frac{\left(p_{\nu1}+p_{\nu2}-p_{rel}\right)^{2}}{2\left(\sigma_{P}^{2}+\sigma_{rel}^{2}\right)}\right]|\nu(\vec{p}_{\nu1})\rangle\langle\nu(\vec{p}_{\nu2})|.\label{eq:NeutrinoDensityMatrix}
\end{equation}
To find the position-space density matrix, project Eq.~\ref{eq:NeutrinoDensityMatrix}
onto position basis states left and right,
\begin{equation}
\rho_{\nu}(y_{1},y_{2})={\cal N}\int dp_{\nu1}dp_{\nu2}\exp\left[-\frac{1}{8}\left(\frac{1}{\sigma_{P}^{2}}+\frac{1}{\sigma_{rel}^{2}}\right)\left(p_{\nu1}-p_{\nu2}\right)^{2}-\frac{\left(p_{\nu1}+p_{\nu2}-p_{rel}\right)^{2}}{2\left(\sigma_{P}^{2}+\sigma_{rel}^{2}\right)}+ip_{\nu1}y_{1}-p_{\nu2}y_{2}\right].
\end{equation}
Switching coordinates to $p_{\pm}=p_{\nu1}\pm p_{\nu2}$ and $y_{\pm}=y_{1}\pm y_{2}$
(including a factor of $\frac{1}{2}$ for the Jacobian of the integral
transformation), 
\begin{equation}
\rho_{\nu}(y_{1},y_{2})={\cal N}\frac{1}{2}\int dp_{-}dp_{+}\exp\left[-\frac{1}{8}\left(\frac{1}{\sigma_{P}^{2}}+\frac{1}{\sigma_{rel}^{2}}\right)p_{-}^{2}-\frac{\left(p_{+}-p_{rel}\right)^{2}}{2\left(\sigma_{P}^{2}+\sigma_{rel}^{2}\right)}+i\left(p_{+}y_{-}+p_{-}y_{+}\right)\right].
\end{equation}
The two $p$ integrals are now just Gaussian Fourier transforms,
\begin{equation}
\rho_{\nu}(y_{1},y_{2})={\cal N}\frac{1}{2}\exp\left[-\frac{1}{2\Delta_{D}^{2}}(y_{1}+y_{2})^{2}-\frac{1}{8\Delta_{OD}^{2}}(y_{1}-y_{2})^{2}-ip_{rel}(y_{1}-y_{2})\right],
\end{equation}
written here in a form to be comparable with Eq. 16 of Ref.~\cite{jones2023width}. From this expression we can read off the diagonal and off-diagonal widths in
position space,
\begin{equation}
\Delta_{D}^{2}=\frac{1}{4}\left(\frac{1}{\sigma_{P}^{2}}+\frac{1}{\sigma_{rel}^{2}}\right),\quad\Delta_{OD}^{2}=\frac{1}{4(\sigma_{P}^{2}+\sigma_{rel}^{2})}.
\end{equation}
The neutrino wave packet width in this formalism is associated with
the value of $\Delta_{OD}$. 
\begin{equation}
\sigma_{\nu,x}\sim\sqrt{\Delta_{OD}}=\frac{1}{2\sqrt{\sigma_{P}^{2}+\sigma_{rel}^{2}}}.
\end{equation}
Here $\sigma_P$ is the momentum uncertainty associated with motion of the atom in the lattice, and $\sigma_{rel}$ is determined by the relative momentum between electron and nucleus in electron capture, as calculated above. It is clear that there is an upper limit to this width, determined by the $\sigma_{rel}$ term. In the case of BeEST ($Z=4$) this corresponds to
\begin{equation}
\sigma_{\nu,x}\leq\frac{a_{0}^{2}m_{e}}{\zeta Z^{2}} = 2.7\,\mathrm{nm}.
\end{equation}
The issues surrounding localization in the material lattice are more subtle.  It is an experimental fact that nuclei must be localized within a lattice plane to a significantly smaller distance scale than the lattice spacing, in order for sharp peaks in X-ray, electron and neutron diffraction experiments to be observable.  The extent to which atoms are delocalized from their average lattice sites is an experimentally accessible quantity, which impacts the strength of lines in X-ray and neutron spectra via the Debye-Waller factor, as discussed in Ref.~\cite{debye1913interferenz}, and in M$\ddot{\mathrm{o}}$ssbauer spectroscopy via the Lamb-M$\ddot{\mathrm{o}}$ssbauer factor, as discussed in Ref.~\cite{Mossbauer1958}.    While broadening of X-ray diffraction peaks could be interpreted classically in terms of a thermal distribution of atomic positions around their lattice sites, connections to the recoilless fraction of absorption events via the M$\ddot{\mathrm{o}}$ssbauer effect make clear that this mean-squared displacement $\langle u^2\rangle $ can also be interpreted as a coherent quantum width, as discussed in Ref.~\cite{Frauenfelder1962}.

The Debye-Waller parameter $B$, defined as
\begin{equation}
    B=8\pi^2 \langle u\rangle ^2
\end{equation} 
has been calculated for tantalum at room temperature~\cite{lottner1979debye} giving mean-square displacement $\langle u^2\rangle\sim4\times10^{-23}~\mathrm{m}^2$. Measurements with X-ray powder diffraction indicate  $\langle u^2\rangle\sim1.5\times10^{-22}~\mathrm{m}^2$~\cite{jiang2003investigation}.  A change by a factor of approximately two in the peak heights between 300~K and 20~K suggests a relatively strong temperature dependence, with $\langle u\rangle^2(20~\mathrm{K})\sim \frac{\langle u\rangle^2(300~\mathrm{K})}{\ln 2}$, approximately $\langle u^2\rangle\sim1\times10^{-22}~\mathrm{m}^2 $.

The Lamb-M$\ddot{\mathrm{o}}$ssbauer factor relates the recoil-free (no phonon) transitions to total number of transitions fraction ($f$) to the mean-square vibrational amplitude of the nucleus in its lattice site $\langle u^ 2\rangle$ via~\cite{Frauenfelder1962}:
\begin{equation}
f=\exp\Big[-\Big(\frac{E_\gamma}{\hbar c}\Big)^2\langle u^2\rangle\Big]
\end{equation}
For the M$\ddot{\mathrm{o}}$ssbauer isotope of tantalum ($^{181}$Ta), the gamma energy ($E_\gamma$) is 6.2~keV. The fraction $f$ depends on temperature, crystal phase, and orientation.  The magnitude of the resonance effect, $\epsilon$, from $^{181}$Ta in a Ta metal source with a Ta metal absorber ($\epsilon\sim f^2$) at room temperature is reported as 2.4\% in Ref.~\cite{shenoy1978mossbauer}. With $f$ in this case at $\sim$15\%,  $\langle u^2 \rangle$ is inferred to be $\sim$2$\times 10^{-21}$~m$^2$. As with the X-ray derived values, this is expected to increase at lower temperatures. From the M$\ddot{\mathrm{o}}$ssbauer effect we thus infer the localization scale in a Ta metal lattice to be around 0.03~nm at room temperature. It will be reduced from this value for the case of cryogenic temperatures in BeEST (e.g. a localization of 0.01 nm is expected for a Lamb-M$\ddot{\mathrm{o}}$ssbauer factor $f$ of 90\%).

Assuming these observables can be considered as a reasonable indication for the quantum mechanical localization of a $^7$Be atom embedded in a tantalum metal lattice, then the center-of-mass localization is more limiting than the $\sigma_{rel}$, giving a distance scale of order 
\begin{equation}
\sigma_{\nu,x}=\frac{1}{\sigma _P}\sim0.01~\mathrm{nm}.
\end{equation}

In summary, we determine that an absolute upper limit on the coherent neutrino wave packet width should be $\sim$2.7~nm, with the actual value depending on details of localization in the atomic lattice, with X-ray and M$\ddot{\mathrm{o}}$ssbauer data in tantalum metal indicating a likely distance scale of $\sim$0.01~nm, a value that is fairly close to the current BeEST limit, as discussed below.

\section{Extracting constraints on the neutrino spatial width from BeEST data}
\label{map_recoil_neutrino}
Along with predicting the outgoing neutrino width following an electron capture decay, a connected issue is the method by which sensitivity to the width can be derived based on measuring the nuclear recoil energy spectrum. 

The authors of Ref. \cite{Smolsky:2024uby} aim to contrast the
approaches advocated by different schools of thought on this matter.
We note that the first approach, assigning equal energy width to neutrino
and recoil, is one we too would subscribe to in the case where the
energy width were limited by intrinsic line width of the decay. However,
since the electron capture lifetime is 53 days, this would imply an
energy width of $10^{-22}\mathrm{\,eV}$. It is well understood that
it is not this width that is encoded in measurement of the recoil
system in electron capture. Instead, this quantity is dictated by homogeneous
broadening due to widths associated with de-excitation processes in
the daughter atom \cite{bambynek1977orbital}, as well as inhomogeneous
broadening associated with solid state effects in materials~\cite{samanta2023material}.

Both the effects of homogeneous broadening due to finite lifetime and inhomogeneous broadening due to the variable environment of the emitter can be incorporated simply into the present formalism.  The energy broadening would appear as a widening of the $\delta$ function in Eq.~\ref{eq:Operator}, as was done in Ref.~\cite{jones2023width}. Due to the very small width expected, this is not understood to limit the neutrino coherent width in this system.  Accounting for the inhomogeneous broadening due to variability of the initial state amounts to taking a weighted sum of reduced density matrices for neutrinos produced from emitters in each possible kind of vacancy, as given in Ref.~\cite{samanta2023material}. Since this represents an additional source of classical uncertainty about the initial state, which is in any case traced out of the final state density matrix incoherently, it will not impact the prediction for the off-diagonal width that dictates neutrino spatial coherence via $\sigma_{\nu,x}$. While neither of these effects will modify the predictions of this calculation concerning the predicted spatially coherent width of the neutrino wave packet, they do broaden the electron capture energy peak, implying that it must be a lower limit that is extracted from BeEST data.  Ref.~\cite{Smolsky:2024uby} recognizes this clearly, and appropriately derives a lower limit on the neutrino width.

The approach we advocate for in assigning a neutrino spatial width limit
is to compare the expected off-diagonal width of the neutrino position-space
density matrix to the on-diagonal width of the recoil momentum-space
density matrix. The latter quantity is what is encoded in the energy
spread measurement that is accessed experimentally by BeEST. Running
through an argument similar to that in the previous section, the momentum
space reduced density matrix for the recoil is
\begin{equation}
\rho_{N'}(p_{1},p_{2})={\cal N}\exp\left[-\frac{1}{2(2\Delta_{OD})^{-2}}(p_{1}+p_{2})^{2}-\frac{1}{2(2\Delta_{D})^{-2}}(p_{1}-p_{2})^{2}\right],
\end{equation}
from which we can extract that
\begin{equation}
\sigma_{N,p'}=\sigma_{\nu,p}=\frac{1}{2\sigma_{N',x}}=\frac{1}{2\sigma_{\nu,x}}.\label{eq:MomentumComp}
\end{equation}


As such, we recommend the following prescription to extract a
lower limit on the coherent width of the neutrino wave packet. First,
note from Eq.~\ref{eq:MomentumComp} that $\sigma_{\nu,p}=\sigma_{N,p'}$.
Intuitively this can be understood by considering that the two particles
recoil against each other in approximately the center of mass with
equal momenta. Then, taking the measured energy spectrum width of
the recoil of $\sigma_{N',E}=$2.9 eV and applying the mapping to
momentum width as described in Ref. \cite{Smolsky:2024uby}, 
\begin{equation}
\sigma_{\nu,p}=\sigma_{N,p'}=\sqrt{m/2E}\sigma_{N',E},
\end{equation}

Finally, applying the Heisenberg uncertainty principle for the neutrino
wave packet, we find
\begin{equation}
\sigma_{\nu,x}=\sigma_{N',x}\geq\frac{\hbar}{2\sigma_{\nu,p}}=6.2\,\mathrm{pm},
\end{equation}
with equality in the case where the wave packet is Gaussian. This is the limit on the spatial neutrino wave packet width from BeEST
data to which our prediction above, $\sigma_{\nu,x}\sim0.01-2.7\,\mathrm{nm}$,
should be consistently compared.

\section{Coherence Effects and the Gallium Anomaly}
\label{gallium_anomaly}

The neutrinos produced in electron capture sources have been detected in radiochemical experiments based on the process $^{71}\mathrm{Ga}(\nu_e,e^-)^{71}\mathrm{Ge}$~\cite{elliott2023gallium}.  These measurements were initially intended as calibration studies for the solar neutrino experiments SAGE~\cite{abdurashitov1996russian} and GALLEX~\cite{anselmann1995first}, with neutrino emission intensities which could in principle be predicted to sub-1\% levels of uncertainty.  The detected neutrino yield was found to be lower than expectations by a ratio R$=0.87\pm0.05$, representing a 2.5$\sigma$ deviation from 1. This ``gallium anomaly'' was subsequently re-interpreted as evidence of novel oscillation physics at short baselines~\cite{bahcall1997gallium}.  The BEST~\cite{barinov2022results} experiment (not to be confused with BeEST, discussed above), was recently performed to re-investigate the gallium anomaly, confirming the effect at the previously reported magnitude and with 4$\sigma$ statistical significance.

The interpretation of the  gallium anomaly in terms of short baseline neutrino oscillations is called into question by strong tensions with other experiments that have not detected anomalies. This has prompted several authors to consider wave packet effects as a possible origin for the gallium anomaly~\cite{farzan2023decoherence,krueger2023decoherence,arguelles2023impact,akhmedov2022damping,jones2022comment}.  In these works, the expected wave packet width is either considered as a free parameter to be fit to data, or else fixed to another physical scale in the problem while neglecting the effects of environmental entanglements that act to localize the neutrino emitter.

The electron capture sources used in radiochemical neutrino experiments to date are $^{51}$Cr and $^{37}$Ar, with anomalies observed using both sources. These two isotopes have electron capture Q-values of 752.39$\pm$0.15~keV and 813.87$\pm$0.20~keV respectively~\cite{be2006table}. $^{51}$Cr also has a 9.91$\pm$0.02\% branching fraction to the first excited state of $^{51}$V, which is accompanied by  a 320.0835$\pm$0.0004~keV $\gamma$ ray as it de-excites back to the ground state. This produces a neutrino with approximately 320~keV below the Q-value in these decays.  For both isotopes, approximately $\sim$90\% of the electron captures are from the K-shell, with $\sim$9\% from the L shell and $\sim$1\% from the M shell. Since the K-shell represents the most spatially localized source where the largest decoherence effects are to be expected, we consider only this subset of decays in this section.

Using the formalism described above, we can estimate the expected wave packet width emerging from the K-shell electron captures.  These numbers can be used to predict the coherence distance of these neutrinos using the formula~\cite{jones2023width}, 
\begin{equation}
    L_{coh}\sim2\sqrt{2}\frac{E_{\nu}^{2}}{\Delta m_{\nu}^{2}}\sigma_{\nu}.
\end{equation}
The $^{51}$Cr source is a solid material, implying a localization width consistent with the lattice delocalization scale, of order 0.01~nm. This gives a coherence distance of 6400~m for the decay to ground state and 2100~m for the excited state decay, assuming the larger of the known $\Delta m^2$ values for active neutrinos ($\Delta m_{32}^2\sim2.5\times 10^{-3}~\mathrm{eV}^2$)~\cite{navas2024review}.  The $^{37}$Ar source, on the other hand, does not involve decays from a solid lattice, and the mean spacing of atoms in the gas is much larger than the atomic size.  As such, the atomically limited width is more appropriate in this case, $\sigma_\nu\sim0.13~\mathrm{nm}$, leading to a coherence distance of 97~km. This is not sufficiently short to produce observable effects in the gallium experiments, which require a coherence distance of between 0.8 and 5~m to match the observed disappearance effect at the 2$\sigma$ confidence level~\cite{farzan2023decoherence}. The 10\% of captures from the higher L and M shells that we have ignored have even longer coherence lengths due to the larger delocalization of the L and M orbitals relative to the K orbitals, so these will also not experience significant decoherence over the relevant experiment baselines.

For larger $\Delta m^2$ values associated with sterile neutrinos, shorter coherence distances would be expected. For values of $\Delta m^2$ of a few electron volts, as indicated by oscillation anomalies~\cite{gariazzo2017updated,moulai2020combining,aartsen2020ev}, we expect the neutrinos produced by $^{51}$Cr to have a coherence length of around 10~m, too large to have significant impact within the BEST detector volume. We estimate that if the wave packet width were were around $5\times10^{-13}~$m then it would impact the BEST experiment's sensitivity to high-$\Delta m^2$ oscillations. In addition, we note that the SAGE anomaly is driven primarily by their $^{37}$Ar data rather than by the $^{51}$Cr data. This is unexpected if neutrino decoherence is contributing to the rate deficit due to the significantly longer coherence distance of $^{37}$Ar compared to $^{51}$Cr. 

Given our predictions for the electron capture neutrino wavepacket width, eV-scale sterile neutrino oscillations augmented by significant wave packet separation do not appear to offer a viable resolution of the gallium anomalies that is consistent with the world's oscillation data.

\section{Conclusion}

We have used a density matrix formalism to consider the relevant distance scales in the electron capture decay of $^7$Be ($e^- + \mathrm{^{7}Be}\rightarrow\mathrm{^{7}Li+\nu_e}$) to predict the width of the emerging neutrino wave packet. We find that the uncertainty associated with the relative momentum between the initial state electron and parent nucleus sets an upper limit at $\sigma_{\nu,x}\leq 2.7$~nm, but a more localizing distance scale is provided by the uncertainty of the decaying emitter position within the tantalum lattice, on a scale of 10~pm. This can be compared to recent experimental results from the BeEST Collaboration ($\sigma_{\nu,x}>6.2$~pm). 

We note that our conclusions have been updated since the first version of this paper appeared on the arXiv, where we claimed a neutrino wave function width prediction of 2.7~nm.  This adjustment in perspective emerges from our recent recognition of the relevance of the Lamb-M$\ddot{\mathrm{o}}$ssbauer and Debye-Waller factors that indicate the localization scales of atoms in solid lattices, which inform the quantum localization scale of $^7$Be atoms within the tantalum matrix relevant for the BeEST experiment.

\section{Acknowledgements}

We thank Kyle Leach, Joe Smolsky and David Moore for fruitful discussions, and we congratulate the BeEST
collaboration for their beautiful experiment. Thanks to Varghese Chirayath, Krishan Mistry, Reagan Miller for their input on this manuscript. BJPJ is supported by
the US Department of Energy under awards DE-SC0019054 and DE-SC0019223. JS is supported by the Department of Energy, Office of Science, under Award No. DE-SC0007859. 

\bibliographystyle{unsrt}
\bibliography{biblio}

\end{document}